\documentstyle[12pt]{article}
\advance\voffset by -1.0in
\advance\hoffset by -0.75in
\newcommand{\Bbb}{\bf}
\newcommand{\Hil}{{\cal H}}
\newcommand{\kay}{{\cal K}}
\newcommand{\ze}{{\Bbb Z}}
\newcommand{\re}{{\Bbb R}}

\newcommand{\reg}[1]{(\ref{#1})}
\begin{document}\title{Intertwiners in Orbifold Conformal Field Theories}
\author{P.S. Montague\\
Department of Mathematics\\
University of Adelaide\\
Adelaide\\
South Australia 5005\\
Australia}
\maketitle
\begin{abstract}
Following on from earlier work relating modules
of meromorphic bosonic conformal field theories
to states representing solutions of certain simple equations
inside the theories, we show, in the context of orbifold theories,
that the intertwiners between twisted sectors are unique and described
explicitly in terms of the states corresponding to the relevant modules.
No explicit knowledge of the structure of the twisted sectors is required.
Further, we propose a general set of sufficiency conditions, illustrated
in the context of a third order no-fixed-point twist
of a lattice theory, for
verifying consistency of arbitrary orbifold models in terms
of the states representing the twisted sectors.
\end{abstract}
\section{Introduction}
In previous papers \cite{PSMreps,PSMtwisunique} we have introduced a
scheme whereby to each representation of a hermitian bosonic
meromorphic conformal field theory (CFT) we assign a certain state,
generically denoted by $P$.  In \cite{PSMreps} a set of equations was
derived which $P$ is required to satisfy, while in
\cite{PSMtwisunique} the general solution to these equations was
found, and in the particular example of $\ze_2$-twisted orbifolds of
the lattice theories this was found to be sufficient to prove the
anticipated ``uniqueness of the twisted representation''.  It is
conjectured in general that all solutions to these equations
correspond to representations (there is a constructive procedure given
in \cite{PSMreps}, though some technicalities remain to be
demonstrated in general), {\em i.e.}  that the correspondence between
representations and solutions of certain equations inside the CFT is
one-to-one. As an immediate application, we see that the calculations
involved in verifying that a $\ze_2$-orbifold of a CFT is itself
consistent as a CFT may be reformulated entirely in terms of the
Hilbert space and vertex operators of the original CFT (specifically
its $\ze_2$-invariant projection), and no explicit realisation
of the twisted sector is necessary, {\em c.f.}
\cite{Hollthesis,DGMtwisted,DGMtriality}.

For orbifolds of order higher than 2, we have also to consider
vertex operators intertwining the various twisted sectors
when discussing the orbifold theory. In this paper, we demonstrate
that similar results hold true, {\em i.e.} that we may completely
define the structure of the orbifold CFT with reference only to states
and vertex operators inside the (twist-invariant projection of the)
original CFT. We show that, given two (twisted) representations corresponding
to states $P_1$ and $P_2$ of a CFT there is at most one intertwiner
corresponding to the action of one on the other, and further we demonstrate
how to calculate the appropriate matrix elements in terms of $P_1$ and
$P_2$. Thus, given that the above mentioned conjecture of a one-to-one
correspondence is true, we can verify completely all of the locality relations
necessary for the orbifold CFT to be consistent entirely in terms of
the $P$'s corresponding to the twisted sectors and the structure of the
original CFT. We stress again that
no explicit knowledge of the twisted sectors is necessary (though it is all
encoded in the $P$'s) and the method is thus of great import when a
geometric interpretation of the orbifold is lacking and it is therefore
unclear how to begin the construction explicitly.
Applications to physically realistic string models and also
to completion of the classification of central charge 24 self-dual
conformal field theories, by explicitly constructing the
possible theories identified by Schellekens
\cite{Schell:Venkov,SchellComplete} as orbifolds
of the existing theories, are immediately apparent. 

As an illustrative example, we consider the no-fixed-point
$\ze_3$-orbifold of a lattice theory
$\Hil(\Lambda)$
discussed in \cite{PSMthird}. The complicated (and incomplete) calculations
therein necessary to verify consistency of the orbifold theory may be
recast entirely in terms of two states $P_1$ and $P_2$ in $\Hil(\Lambda)$
corresponding to the two twisted sectors (we have two non-trivial
conjugacy classes in the $\ze_3$ group of automorphisms $\langle\theta
\rangle$). In \cite{PSMthird}, an ansatz
was made for the form of the intertwiner between the two twisted sectors
and the coefficients explicitly evaluated (recovering results of Gato in
\cite{Gato:tvos}). As we now see, this is not necessary -- no assumptions
need be made. Also, in this case, an explicit realisation of the twisted
sectors is known, and we do not even require our conjecture about the
one-to-one correspondence between solutions $P$ of the
equations of \cite{PSMreps,PSMtwisunique} and representations to hold true.
Consistency of the $\ze_3$-orbifold may hence be verified, and further we
demonstrate that much of the calculation of \cite{PSMthird}
is unnecessary due to a relation
between $P_1$ and $P_2$ which generalises to arbitrary orbifold
theories.

The layout of the paper is as follows.

In section \ref{P} we summarise and generalise to distinct,
non-quasi-primary and non-real
representation states the results of \cite{PSMreps} and
\cite{PSMtwisunique}.

In section \ref{Q}, we extend this work to consider intertwiners between
distinct representations, while in section \ref{eg} we illustrate our
discussion first with a trivial application to the Heisenberg algebra
before considering the more sophisticated
example of the no-fixed-point $\ze_3$-orbifold
of $\Hil(\Lambda)$.

In the appendix, we demonstrate the rather surprising result
that an apparently natural (and considerably more amenable to
calculation) ansatz for the form of the twisted sector-twisted
sector intertwiner necessarily fails in general.
\section{Construction of the state $P$}
\label{P}
Let us first establish our notation. We define a conformal field theory
(strictly a bosonic, hermitian, meromorphic conformal field theory)
to consist of a Hilbert space $\Hil$, two fixed states $|0\rangle$, $\psi_L\in
\Hil$,
 and a set $\cal V$ of ``vertex operators'',
{\em i.e.} linear operators
$V(\psi,z):\Hil\rightarrow\Hil$, $\psi\in\Hil$ parameterized by a complex
parameter $z$ such that $V(\psi_1,z_1)V(\psi_2,z_2)\cdots$ makes
sense for $|z_1|>|z_2|>\cdots$,
\begin{equation}
\label{create}
V(\psi,z)|0\rangle=e^{zL_{-1}}\psi\,,
\end{equation}
\begin{equation}
\label{locality}
V(\psi,z)V(\phi,w)=V(\phi,w)V(\psi,z)
\end{equation}
(the so-called ``locality'' relation -- it
holds in the sense that appropriate analytic continuations of matrix
elements of either side agree) and
\begin{equation}
V(\psi_L,z)\equiv\sum_{n\in\ze}L_nz^{-n-2},,
\end{equation}
where
\begin{equation}
[L_m,L_n]=(m-n)L_{m+n}+{c\over 12}m(m^2-1)\delta_{m,-n}\,,
\end{equation}
(the constant $c$ is the ``central charge'' of the theory),
with ${L_n}^\dagger=L_{-n}$.
See \cite{DGMtriality} for a full discussion of this definition (as well
as the more technical axioms omitted here). We will use the result
\begin{equation}
\label{fred}
\left[ L_{-1},V(\psi,z)\right]={d\over dz}V(\psi,z)\,.
\end{equation}

We define a representation (which we will take to be real and hermitian --
see \cite{DGMtriality})
of this theory to consist of a Hilbert space
$\kay$ and a set of linear (vertex) operators $U(\psi,z):\kay\rightarrow
\kay$, $\psi\in\Hil$, such that ``duality'' holds, {\em i.e.}
\begin{equation}
\label{duality}
U(\psi,z)U(\phi,w)=U(V(\psi,z-w)\phi,w)\,,
\end{equation}
and also $U(|0\rangle)\equiv 1$. Note that a relation identical to 
\reg{duality} is satisfied by the $V$'s as a consequence of the above axioms
\cite{PGmer,DGMtriality}. This representation is said to be meromorphic if
matrix elements of the $U$'s are meromorphic functions of the complex
arguments.

We also define vertex operators $W$ and the conjugate $\overline W$ by
\begin{equation}
W(\chi,z)\psi=e^{zL_{-1}}U(\psi,-z)\chi
\label{Wdef}
\end{equation}
and
\begin{equation}
\label{conj}
\overline W(\chi,z)=W\left(e^{z^\ast L_1}{z^\ast}^{-2L_0}
\overline\chi,{1\over z^\ast}\right)^\dagger\,,
\end{equation}
where $\chi\mapsto\overline\chi$ is a certain antilinear involution
(see \cite{DGMtriality} for details). [Note that, in
general, $\overline\chi$ may lie in a distinct representation
to $\chi$. In particular, in the case of the twisted representations
which
we will discuss below, if $\chi$ lies in a $g$-twisted representation
than its conjugate $\overline\chi$ lies in a $g^{-1}$-twisted
representation.] 
There is a similar involution on $\Hil$, and the relation
\begin{equation}
\label{herm}
Y(\overline\psi,z^\ast)=Y\left(e^{zL_1}z^{-2L_0}\psi,1/z\right)^\dagger
\end{equation}
holds with $Y$ representing either $V$ or $U$.
Note that the operators
$W$ satisfy the ``intertwining relation''
\begin{equation}
\label{freddie}
W(\chi,z)V(\psi,w)=U(\psi,w)W(\chi,z)\,,
\end{equation}
with a similar relation holding for $\overline W$ (as a consequence of
($\ref{conj}$) and ($\ref{herm}$).
[Note that this is simply a relation of the form ($\ref{locality}$)
again. This will be a general feature of the following. Any
relation of
the form ($\ref{locality}$) or ($\ref{duality}$) which makes sense
({\em i.e.} in which the states are matched to appropriate
vertex operators) holds.]

Consider fixed states $\chi$ and $\rho$ in the representation.
Note that we no longer
assume that $\chi$ or $\rho$
are equal, real and quasi-primary, as we did in \cite{PSMreps}.
Then, for $\psi_j\in\Hil$ for $1\leq j\leq n$, the matrix element
\begin{eqnarray}
\langle\chi|U(\psi_1,z_1)&\ldots&U(\psi_n,z_n)|\rho\rangle\nonumber\\
&=&
\langle\chi|U(\psi_1,z_1)\ldots U(\psi_{n-1},z_{n-1})e^{z_nL_{-1}}
W(\rho,-z_n)|\psi_n\rangle\ \ \hbox{by (\ref{Wdef})}\nonumber\\
&=&\langle\chi|e^{z_nL_{-1}}U(\psi_1,z_1-z_n)\ldots U(\psi_{n-1},
z_{n-1}-z_n)W(\rho,-z_n)|\psi_n\rangle\ \ \hbox{by (\ref{fred})}\nonumber\\
&=&\langle\chi|e^{z_nL_{-1}}W(\rho,-z_n)V(\psi_1,z_1-z_n)\ldots V(\psi_{n-1},
z_{n-1}-z_n)|\psi_n\rangle\ \ \hbox{by (\ref{freddie})}\nonumber\\
&\equiv&\langle P(\chi,\rho;{z_n}^\ast)|V(\psi_1,z_1-z_n)\ldots V(\psi_{n-1},
z_{n-1}-z_n)|\psi_n\rangle\,,
\label{trev}
\end{eqnarray}
where
\begin{eqnarray}
\label{Pdef}
P(\chi,\rho;z)&=&W(\rho,-z^\ast)^\dagger e^{zL_1}|\chi\rangle\nonumber\\
&=&\overline W\left(e^{-zL_1}z^{-2L_0}\overline\rho,-{1\over z}\right)
e^{zL_1}|\chi\rangle\,,
\end{eqnarray}
or, as in \cite{PSMtwisunique},
\begin{equation}
\label{def}
\langle P(\chi,\rho;z^\ast)|\psi\rangle=\langle\chi|U(\psi,z)|\rho\rangle\,.
\end{equation}
We can easily modify the equations for $P$ derived in \cite{PSMreps}
to this more general case.
Rather than proceed as in \cite{PSMreps}, we can simply check the relations
directly from the definition ($\ref{def}$).
\begin{itemize}
\item $\langle P(\chi,\rho;z^\ast)|0\rangle=\langle\chi|\rho\rangle$.
\item $\langle P(\chi,\rho;z^\ast)|=
\langle P(\chi,\rho;w^\ast)|e^{(z-w)L_{-1}}$: This follows
immediately using ($\ref{fred}$).
\item $\langle\overline{P(\rho,\chi;z)}|=
\langle P(\chi,\rho;1/z^\ast)|e^{zL_1}z^{-2L_0}$:
If we take the inner product with a state $\psi$,
the right hand side becomes
\begin{equation}
\langle\chi|U\left(e^{zL_1}z^{-2L_0}\psi,1/z\right)|\rho\rangle
=\langle\rho|U(\overline\psi,z^\ast)|\chi\rangle^\ast\,,
\end{equation}
using ($\ref{herm}$). Using the definition again, and the result
\begin{equation}
\label{hermint}
\langle \psi|\overline\phi\rangle=\langle\phi|\overline\psi\rangle
\end{equation}
(see \cite{DGMtriality}) gives us the left hand side.
\item $\overline{P(\chi,\rho;z)}=P(\overline\chi,\overline\rho;-z^\ast)$:
This is easily verified using the
``reality'' of the representation \cite{DGMtriality}
\begin{equation}
\langle\eta|U(\psi,z)|\phi\rangle^\ast=
\langle\overline\eta|U(\overline\psi,-z)|\overline\phi\rangle\,,
\end{equation}
together with ($\ref{hermint}$).
\end{itemize}
The claim made in \cite{PSMreps}, and justified up to technicalities
regarding the rigorous definition of the Hilbert space for the
representation, is that these equations on $P$ (in
the case $\chi=\rho=\overline\chi$) are also {\bf sufficient}
for it to define a representation (the calculations of \cite{PSMreps}
show that using ($\ref{trev}$) as a definition
of the vertex operators $U$
satisfies the appropriate properties for a representation, provided
the Hilbert space can be understood). The obvious relation to work of Zhu
\cite{Zhu} remains to be understood precisely.

In \cite{PSMtwisunique}, the general solution to the above equations
was found, and further it was shown that a necessary (and sufficient)
condition for the state $\chi$ to be a ground state in the abstract
representation defined via ($\ref{trev}$) is that $P\equiv P(1)$
be orthogonal to what Zhu denotes by $O(\Hil)$ -- a space of states
whose zero modes automatically annihilate the ground states of all
representations of the CFT $\Hil$.
\section{Construction of intertwiners between non-trivial
representations}
\label{Q}
Suppose we have two representations $\kay_1$ and
$\kay_2$ of a CFT $\Hil$.
We wish to construct the intertwiner representing the action
of $\kay_1$ on $\kay_2$, i.e. a set of vertex operators corresponding
to states in $\kay_1$ acting on $\kay_2$. Let us assume that
the two representations fuse to give only one representation.
The canonical example which
we wish to keep in mind is that in which we are considering
orbifolding $\Hil$ with respect to some finite group $G$ (which we
take to be abelian for simplicity), and $\kay_1$ and $\kay_2$ are
$g_1$ and $g_2$-twisted representations respectively for certain
group elements $g_1$ and $g_2$. (See \cite{DLMZhu}
for definitions
and existence theorems on such twisted sectors.)
In this case, the intertwiner maps us to a $g_1g_2$-twisted
representation.

The case discussed in the previous section is that in which the
intertwiner maps back to $\Hil$ (the intertwining vertex
operators are then simply the $\overline W$'s), {\em i.e.} in the case of the
orbifold theory where $g_1={g_2}^{-1}$.  We consider here the more
general case in which the image space is a non-adjoint representation
of $\Hil$. Let us denote it $\kay_3$.

Choose fixed quasi-primary states $\chi_1$ and $\chi_2$
in $\kay_1$ and $\kay_2$ respectively (typically we will
take these to lie in the ground state of the representation --
note that we cannot choose them to be real, however,
due to the comment following equation ($\ref{conj}$)),
and of respective conformal weights $h_1$ and $h_2$.
Suppose the CFT $\Hil$ is
represented on $\kay_1$, $\kay_2$ and $\kay_3$ by vertex operators
$U_1$, $U_2$
and $U_3$ respectively. We wish to define an intertwining
operator $\hat W(\rho,z):\kay_2\rightarrow\kay_3$ for $\rho\in\kay_1$.
Using the fact that the representations are irreducible representations
of $\Hil$, all we need to define is the state
\begin{equation}
\hat W\left(U_1(\psi_1,w_1)\chi_1,z\right)U_2(\psi_2,w_2)\chi_2\,,
\end{equation}
for all $\psi_1$, $\psi_2\in\Hil$. As argued in \cite{PSMthird},
this must be given (if we are to have a consistent CFT, {\em i.e.}
if appropriate intertwining relations are to hold) by
\begin{equation}
U_3(\psi_2,w_2)U_3(\psi_1,w_1+z)\hat W(\chi_1,z)\chi_2\,.
\end{equation}
Hence, we can identify the required intertwiner once we know
$R(\chi_1,\chi_2;z)\equiv\hat W(\chi_1,z)\chi_2$. In the appendix, we demonstrate that an obvious
ansatz at this point cannot work in general. Let us proceed instead
as follows.

Expand $R(\chi_1,\chi_2;z)$ as
\begin{equation}
\label{Rexp}
R(\chi_1,\chi_2;z)=\sum_{n\geq 0}\eta_nz^{n-\delta}\,,
\end{equation}
$\eta_n\in\kay_3$ (and suppressing the dependence on $\chi_1$
and $\chi_2$ for the sake
of notational compactness), for some shift $\delta$ with $\eta_0\neq 0$. The
conformal weight of $\eta_n$ is $n+h_3$, for
some constant $h_3$. Denote $\eta_0$ by
$\chi_3$. (Note that $\delta=h_1+h_2-h_3$.) [Typically, $\chi_3$ will
lie in the ground state of $\kay_3$, though we do not know in general
when this must be so and so cannot assume it. This is one of the
crucial problems that we have to overcome in the following.]

In order to map $R(\chi_1,\chi_2;z)$ to a state inside the original CFT $\Hil$,
we act with the conjugate intertwining operator $\overline W_3$
({\em c.f.} ($\ref{conj}$)), where $W_3$ is the
intertwiner corresponding to $U_3$ as in
($\ref{Wdef}$). Following ($\ref{Pdef}$),
we consider the state
\begin{equation}
\label{Qdef}
Q(\chi_1,\chi_2;w,z)\equiv \overline W_3\left(e^{-wL_1}w^{-2L_0}\overline
{\chi_3},-{1\over w}\right)e^{wL_1}R(\chi_1,\chi_2;z)
\end{equation}
in $\Hil$. (Note that the state $\chi_3$ is, at this point, still
unknown.)
Another way to look at this is to consider the following matrix
elements which, by irreducibility of $\kay_3$ as a $\Hil$ representation,
are clearly sufficient, if known for
all $\psi\in\Hil$, to determine $R(\chi_1,\chi_2;z)$ completely.
\begin{eqnarray}
&&\langle\chi_3|U_3(\psi,w)\hat W(\chi_1,z)|\chi_2\rangle\nonumber\\
&=&\langle\chi_2|{\hat W}^\dagger(\chi_1,z)U_3\left(
e^{w^\ast L_1}{w^\ast}^{-2L_0}\overline\psi,{1\over w^\ast}\right)
|\chi_3\rangle^\ast\ \ \ \ \ \hbox{by ($\ref{herm}$)}\nonumber\\
&=&\langle\chi_2|{\hat W}^\dagger(\chi_1,z)e^{{1\over w^\ast}L_{-1}}
W_3\left(\chi_3,-{1\over w^\ast}\right)e^{w^\ast L_1}{w^\ast}^{-2L_0}
|\overline\psi\rangle^\ast\ \ \ \ \ 
\hbox{{\em c.f.} ($\ref{Wdef}$)}\nonumber\\
&=&\langle\overline\psi|w^{-2L_0}e^{wL_{-1}}
\overline W_3\left(e^{-{1\over w}L_1}w^{2L_0}\overline{\chi_3},-w\right)
e^{{1\over w}L_1}W(\chi_1,z)|\chi_2\rangle
\ \ \ \ \ \hbox{by ($\ref{conj}$)}\nonumber\\
&=&\langle\overline\psi|w^{-2L_0}e^{wL_{-1}}|Q(\chi_1,\chi_2;1/w,z)\rangle\,.
\label{21}
\end{eqnarray}
[The expression ($\ref{Qdef}$) may be simplified slightly by noting that
$\chi_3$ is necessarily quasi-primary (in fact, it must be primary),
{\em i.e.}
\begin{eqnarray}
\label{t22}
Q(\chi_1,\chi_2;w,z)&\equiv&\overline W_3\left(w^{-2L_0}\overline
{\chi_3},-{1\over w}\right)e^{wL_1}\hat W(\chi_1,z)\chi_2\nonumber\\
&=&w^{-2h_3}\left({z\over 1-wz}\right)^{2h_1}\overline{W_3}\left(
\overline{\chi_3},-{1\over w}\right)\hat W\left(\chi_1,{z\over 1-wz}\right)\chi_2\,,
\end{eqnarray}
using the $su(1,1)$ transformation properties of primary fields
from {\em e.g.} \cite{PGmer}.]

Hence, knowledge of the state $Q(\chi_1,\chi_2;w,z)\in\Hil$ is sufficient to determine
the intertwining operators $\hat W(\rho,z)$. But we can say much more.
Rather than just treating $Q$ as another state in $\Hil$ to be determined
by equations analogous to those we have to solve for the $P$'s, we shall
show that $Q$ is determined uniquely
in terms of the $P$'s corresponding to
the representations $\kay_1$ and $\kay_2$.

Inserting the expansion ($\ref{Rexp}$) of $R(\chi_1,\chi_2;z)$ in ($\ref{Qdef}$),
we have
\begin{equation}
Q(\chi_1,\chi_2;w,z)=\sum_{n\geq 0}z^{n-\delta}P(\eta_n,\chi_3;w)\,.
\end{equation}

Let us first use this to simplify ($\ref{21}$). We obtain, using the equations
satisfied by the $P$'s described in the previous section,
\begin{eqnarray}
\label{expm}
\langle\chi_3|U_3(\psi,w)\hat W(\chi_1,z)|\chi_2\rangle&=&
\sum_{n\geq 0}z^{n-\delta}\langle\overline\psi|
w^{-2L_0}e^{wL_{-1}}|P(\eta_n,\chi_3;1/w)\rangle\nonumber\\
&=&\sum_{n\geq 0}z^{n-\delta}\langle\overline\psi|\overline{P(\chi_3,\eta_n;w^\ast)}
\rangle\nonumber\\
&=&\sum_{n\geq 0}z^{n-\delta}\langle P(\chi_3,\eta_n;w^\ast)|\psi\rangle\,,
\end{eqnarray}
by ($\ref{hermint}$) (in fact, obvious from the definition of the
$P$'s).

Consider the matrix elements
\begin{equation}
\label{matel}
\langle R(\chi_1,\chi_2;d^\ast)|U_3(\psi,w)|R(\chi_1,\chi_2;z)\rangle\,,
\end{equation}
for arbitrary states $\psi\in\Hil$. By the series expansion of $R$
and the definition of $P$, this is simply
\begin{equation}
\sum_{m,n\geq 0}d^{m-\delta}z^{n-\delta}
\langle\eta_m|U_3(\psi,w)|\eta_n\rangle=\sum_{m,n\geq 0}
d^{m-\delta}z^{n-\delta}
\langle P(\eta_m,\eta_n;w^\ast)|\psi\rangle\,.
\label{series}
\end{equation}
On the other hand, ($\ref{matel}$) is
\begin{eqnarray}
\label{t27}
&&\langle\chi_2|\hat W(\chi_1,d^\ast)^\dagger U_3(\psi,w)\hat W(\chi_1,z)
|\chi_2\rangle\nonumber\\
&=&d^{-2h_1}\langle\chi_2|\overline{\hat W}\left(\overline\chi_1,
{1\over d}\right)\hat W(\chi_1,z)U_2(\psi,w)|\chi_2\rangle\,,
\end{eqnarray}
by an appropriate intertwining relation and the analog
of ($\ref{conj}$),
\begin{equation}
=d^{-2h_1}\langle\chi_2|U_2\left(\overline{W_1}\left(
\overline{\chi_1},{1\over d}-z\right)\chi_1,z\right)U_2(\psi,w)
|\chi_2\rangle\,,
\end{equation}
by an appropriate analog of ($\ref{duality}$),
\begin{equation}
=(1-dz)^{-2h_1}\langle\chi_2|U_2(\psi,w)U_2\left(
P\left(\chi_1,\chi_1;{d\over dz-1}\right),z\right)|\chi_2\rangle\,,
\end{equation}
by the definition of $P$,
\begin{equation}
=(1-dz)^{-2h_1}\langle P(\chi_2,\chi_2;z^\ast)|V(\psi,w-z)
|P(\chi_1,\chi_1;d/(dz-1))\rangle\,,
\label{answer}
\end{equation}
by ($\ref{trev}$).
Thus we may evaluate ($\ref{series}$) in terms of the known $P$'s
corresponding to the representations $\kay_1$ and
$\kay_2$, {\em i.e.} we can, in principle, calculate $P(\eta_m,\eta_n;z)$.
$Q(\chi_1,\chi_2;w,z)$ itself is simply a series in $P(\eta_n,\eta_0;w)$, and thus
is fixed. Alternatively, we can regard the matrix element on the left
hand side of ($\ref{21}$) as being given by the $d^{-\delta}$
term in the expansion
of ($\ref{answer}$) about $d=0$ (obvious from the definitions without resorting
to the expansions in terms of states $\eta_n$, but the approach in terms
of the $\eta_n$'s bears a closer correspondence to the earlier work), or rather of
\begin{equation}
\label{expn}
z^{-h_\psi}z^{-\delta}\langle P(\chi_2,\chi_2;1)|V\left(\psi,{w\over z}-1\right)
|P(\chi_1,\chi_1;d)\rangle
\,,
\end{equation}
where $h_\psi$ is the conformal weight
of $\psi$.
We conclude therefore that the intertwiner representing
the action of $\kay_1$ on $\kay_2$ is determined. The only ambiguity
is in identifying the state $\chi_3$ in the representation $\kay_3$.
If no explicit structure for $\kay_3$ is known, it may simply
be built up from the state $\chi_3$, using our knowledge of
$P(\chi_3,\chi_3;z)$ and ($\ref{trev}$)
via the arguments of \cite{PSMreps}. On the other hand, if
we already have an explicit realization of $\kay_3$, we can
typically identify $\chi_3$ by the way in which the Hilbert space
defined by ($\ref{trev}$) is built up from it ({\em e.g.} at the
simplest level, we can easily calculate its conformal weight). In typical
applications, $\chi_3$ will simply lie in the ground state of the
target representation. Also note that, in cases where we have an explicit
presentation of $\kay_3$, it is often easier to use the above as a means
of defining $R(\chi_1,\chi_2;z)$ rather than $Q(\chi_1,\chi_2;w,z)$, which
is, in some sense, a more fundamental state and can be read off from
comparing the left hand side of ($\ref{expm}$)
to the leading term in the expansion of ($\ref{expn}$).

As
a final comment in this section,
it is interesting to note, from comparing the definition of $P$
with the form of ($\ref{answer}$), that the intertwiner
between (non-trivial) modules
is described entirely by the state
\begin{equation}
P\left(P(\chi_2,\chi_2;1),P(\chi_1,\chi_1,w);z\right)\,.
\end{equation}
\section{Examples and Application to Orbifolds}
\label{eg}
As an illustration of the above points, let us first consider the trivial
example of representations of the Heisenberg algebra. As usual, we take
this to be built up from a vacuum $|0\rangle$ by the action
of creation and annihilation operators satisfying
\begin{eqnarray}
\ [a_m,a_n]&=&m\delta_{m,-n}\ \ \ \hbox{for $m$, $n\in\ze$}\nonumber\\
a_n|0\rangle&=&0\ \ \ \hbox{for $n\geq 0$}\,.
\end{eqnarray}
[As is well known,
appropriate vertex operators and an appropriate conjugation can be defined
to make this a bosonic, hermitian, meromorphic conformal field theory of
unit central charge.] Representations, labelled $\kay_\lambda$,
are built up by an isomorphic set of
operators (we will identify them) acting on a ``momentum state''
$|\lambda\rangle$ of conformal weight
${1\over 2}\lambda^2$, $\lambda\in\re$, satisfying
\begin{eqnarray}
a_n|\lambda\rangle=0\ \ \ \hbox{for $n>0$}\nonumber\\
a_0|\lambda\rangle=\lambda|\lambda\rangle\,.
\end{eqnarray}
[Again, appropriate vertex operators may easily be defined.]
We can show that
\begin{equation}
P(|\lambda\rangle,|\mu\rangle;z)=\delta_{\lambda\mu}e^{-\lambda
\sum_{n>0}(-1)^n{a_{-n}\over n}z^{-n}}|0\rangle\,,
\end{equation}
either from the explicit form of the vertex operators defining the
representation or, more appropriately for the point of view taken in this
paper,  from solving the appropriate equations for $P$
as in \cite{PSMtwisunique}.

Suppose that we wished to consider the intertwiner representing the action
of $\kay_\lambda$ on $\kay_\mu$. From the known structure of the
representations, the fusion results in the representation $\kay_{\lambda+\mu}$
and the state $\chi_3$ in this instance is simply proportional to
the ground state $|\lambda+\mu\rangle$ (we denote
the constant of proportionality
$\epsilon(\lambda,\mu)$ -- see {\em e.g.}
\cite{DGMtwisted} for an explanation
of such factors). Consider a typical matrix element involving the
intertwiner. Suppose we wish to evaluate
\begin{equation}
\langle\lambda+\mu|\epsilon(\lambda,\mu)^\ast U(a_{-1}|0\rangle,w)
W(|\lambda\rangle,z)|\mu\rangle\,,
\end{equation}
where we have dropped suffices from vertex operators for the sake of clarity.
Using the known structure of the vertex operators, this turns out to be
\begin{equation}
\label{magh}
z^{\lambda\mu}\left({\lambda\over w-z}+{\mu\over w}\right)\,.
\end{equation}
Alternatively, by the above, we may calculate this without any direct
knowledge of the structure of the representations or their vertex operators,
merely of the underlying Heisenberg algebra.
We simply evaluate
\begin{equation}
\langle P(|\mu\rangle,|\mu\rangle;z^\ast)|V(a_{-1}|0\rangle,w-z)|
P(|\lambda\rangle,|\lambda\rangle;d/(dz-1))\rangle\,.
\end{equation}
This gives
\begin{equation}
(dz)^{\lambda\mu}\left({\lambda\over w-z}\left({1-dz\over 1-dw}\right)
+{\mu\over w}\right)\,.
\end{equation}
Looking at the leading term in the expansion in $d$ gives ($\ref{magh}$)
(and also shows that the conformal weight of the state ``$\chi_3$''
is ${1\over 2}(\lambda+\mu)^2$, from the leading power of $d$ ({\em i.e.}
$\delta=-\lambda\mu$).

Note that an alternative trivial means of directly calculating the conformal
weight of the state $\chi_3$ in general (and also the norms
of the states $\eta_n$ in the expansion of $R$) is to simply use the
main result of the preceding section with $\psi=|0\rangle$, {\em i.e.}
the identity
\begin{equation}
\sum_{n\geq 0}\alpha^{n-\delta}||\eta_n||^2=(1-\alpha)^{-2h_1}
\langle P(\chi_2,\chi_2;1)|P(\chi_1,\chi_1;\alpha/(\alpha-1))\rangle\,,
\end{equation}
(setting $\alpha=dz$ and using the fact that $P(z)=z^{-L_0}P(1)$).

Before proceeding to the third order orbifold example, some
general comments are in order. Firstly, we note trivially that
the last two of the four equations for $P$ derived in section
2 give
\begin{equation}
\label{Pred}
\langle P(\overline\chi,\overline\rho;-z^\ast)|=
\langle P(\chi,\rho,{1\over z^\ast})|e^{zL_1}z^{-2L_0}\,.
\end{equation}
Suppose that we are orbifolding a CFT $\Hil$ with respect to some
discrete group $G$ with $N$ non-trivial conjugacy classes. We
obtain one twisted sector for each of these classes \cite{Hollthesis}.
Let us choose states $\chi_i$, $1\leq i\leq N$, in the ground
states of the sector twisted by a group element $g_i$. Further,
choose these states so that $\chi_i=\overline{\chi_{j(i)}}$ where
${g_{j(i)}}^{-1}$
lies in the conjugacy class of $g_i$. Set $P_i(z)=P(\chi_i,\chi_i,z)$.
Then, $P_{j(i)}(z)$ is determined in terms of $P_i(z)$ by
($\ref{Pred}$).

We will argue below, at least in the case $G=\ze_3$,
that once the $P_i$'s are specified we need only consider the
matrix elements
\begin{equation}
\langle P_i(1)|V(\psi,w)|P_j(z)\rangle\,,
\end{equation}
for all $\psi\in\Hil$ and all $1\leq i\leq j\leq N$ in order to specify
the orbifold theory (and check its consistency) completely. By the above
observation, we may restrict $i$ and $j$ to a certain subset of $\{1,2,
\ldots,N\}$ ({\em i.e.} a minimal subset $S$ such that $S\cup
j(S)=\{1,2,\ldots,N\}$).

Consider then the case of orbifolding with respect to a
no-fixed-point third order lattice automorphism the conformal field
theory corresponding to an even self-dual lattice $\Lambda$.
In this case, there are two non-trivial conjugacy classes (twisted
sectors).

In order to be more specific, let us give an explicit realisation of
the theory. [See \cite{PSMthird} for full details.]  For notational
convenience, we rewrite the even self-dual lattice admitting a third
order no-fixed-point automorphism in terms of a complex lattice $\Lambda$
over
the ring of Eisenstein integers and of (complex) dimension $d$ (a
multiple of 4).
The CFT (before orbifolding) has Hilbert space $\Hil(\Lambda)$
built up from
``momentum'' states
$|\lambda\rangle$, $\lambda\in\Lambda$,
by the action of creation and annihilation
operators $b_n^i$ and $\overline b_n^i$, $n\in\ze-\{0\}$, $1\leq i\leq d$,
satisfying
\begin{eqnarray}
\ [b_m^i,\overline b_n^j]=m\delta_{m,-n}\delta^{ij}\,;\qquad
[b_m^i,b_n^j]=[\overline b_m^i,\overline b_n^j]=0\nonumber\\
b^i_m|\lambda\rangle=\overline b^j_n|\lambda\rangle=0\ \ 
\hbox{\rm for}\ m,\ n>0\,.
\end{eqnarray}
Appropriate vertex operators and conjugation may be defined to make this a
hermitian, meromorphic, bosonic conformal field theory of central
charge $2d$ (which we also denote $\Hil(\Lambda)$).
It admits an obvious third order automorphism $\theta$ induced by
the third order automorphism $\lambda\mapsto\omega\lambda$,
$\omega=e^{2\pi i/3}$, on the lattice.
We orbifold with respect to the action of this $\ze_3$.

The two twisted sectors have identical Hilbert spaces. In order
to construct them, we introduce a set of ``twisted'' oscillators
$c^i_r$, $1\leq i\leq d$, $r\in\ze\pm{1\over 3}$,
satisfying
\begin{equation}
\ [c^i_r,c^j_s]=r\delta_{r,-s}\delta^{ij}\,.
\end{equation}
Then we construct the Hilbert space $\Hil_1$
built up by the action of these
on a ground state space ${\cal G}$ annihilated by $c^i_r$ for
$r>0$. [${\cal G}$ is essentially a representation space
for a set of cocycle operators required for consistency of the
theory -- see \cite{PSMthird} for more details.] Define a second
space $\Hil_2$ built up using an isomorphic set of oscillators
$\overline c^i_r$. [Note that, though one of the key points
of this paper is that such explicit constructions are, in
general, unnecessary, we wish to relate to earlier work
and some details of the twisted sectors of the orbifold theory
are required.]

$\Hil_1$ and $\Hil_2$ can be given the structure of non-meromorphic
modules for the CFT $\Hil(\Lambda)$. [In defining the action of the
CFT on the modules, there is essentially a choice in associating
states composed of $b$ oscillators with states built up
from twisted oscillators graded by $\ze+{1\over 3}$ (and those
composed of $\overline b$ oscillators with states built up from
twisted oscillators graded by $\ze-{1\over 3}$)  or
with states built up from twisted oscillators graded by
$\ze-{1\over 3}$ (and conversely for $\overline b$) --
we make opposite choices for $\Hil_1$ and $\Hil_2$.]
The ground state ${\cal G}$ turns out (in both modules)
to have conformal weight $d/9$. The oscillators $c^i_r$
(or $\overline c^i_r$) reduce the conformal weight by $r$ as usual.
Let $\Hil_1^0$, $\Hil_2^0$ be the projections
of the modules on to states of integral conformal weight (we must
take $d$ to be a multiple of 12).
It is easily shown that these form {\em irreducible meromorphic}
modules for the sub-CFT $\Hil(\Lambda)^0$ given by the $\theta=1$
sector of $\Hil(\Lambda)$.

The hypothesis is then that $\widetilde\Hil(\Lambda)\equiv
\Hil(\Lambda)^0\oplus\Hil_1^o\oplus
\Hil_2^0$ can be given the structure of a hermitian, meromorphic,
bosonic conformal field theory. What we have shown in the
preceding section is that there is at most one way in which this
may be achieved, {\em i.e.} the intertwiner between the
two twisted sectors required to complete the structure to
a CFT can be defined in only one way in terms of the $P$'s
corresponding to the modules $\Hil_1^0$ and $\Hil_2^0$.
Further, we have shown how the matrix elements of the intertwiner
may be evaluated.

In \cite{PSMthird}, we conjectured an ansatz for the form
of the states in $\Hil_2^0$ given by the action of the intertwiner
corresponding to a state $\chi$ in the ground state of $\Hil_1^0$
on itself, {\em i.e.} the state $R(\chi,\chi;z)$ in the notation
of the previous section, for $d$ a multiple of 36 (so that
${\cal G}\subset\Hil_{1,2}^0$). This is seen to be no longer
necessary, and may be checked by verifying the equality of
($\ref{answer}$) with $\chi_1=\chi_2=\chi$ and $\langle R(\chi,\chi;d^\ast)|U_2(\psi,w)
|R(\chi,\chi;z\rangle$, where the vertex operators $U_2$
give the representation of $\Hil(\Lambda)^0$ on
$\Hil_2^0$, for all $\psi\in\Hil(\Lambda)^0$.
[Note that due to a slight change of notation here,
the state ``$\chi_3$'' lies in $\Hil_2$, and so the vertex operators
labelled $U_3$, $W_3$ previously are here denoted $U_2$, $W_2$.]

In particular, it was assumed that the ground state of $\Hil_2^0$
occurred in the expansion of $R(\chi,\chi;z)$, {\em i.e.} that
the coefficient $\delta$ of the previous section
is given by the conformal weight, $d/9$, of the twisted sector
ground states. This may be easily checked, as remarked above,
by simply evaluating $\langle P(\chi,\chi;1)|P(\chi,\chi;d)\rangle$
and expanding around $d=0$.

It was observed in \cite{PSMthird} that a total of 18 locality
relations need to be checked in order to verify that $\widetilde\Hil(\Lambda)$
is consistent as a CFT.
Most of these follow from locality in $\Hil(\Lambda)$, the fact
that $\Hil_{1,2}^0$ form meromorphic irreducible representations
of $\Hil(\Lambda)^0$ and suitable definitions of conjugation
in the twisted sectors.
There remain to be proved four relations, which we now
translate into our current notation.
\begin{enumerate}
\item $W_1(\chi,z)\overline{W_2}(\chi,w)=W_1(\chi,w)
\overline{W_2}(\chi,z)$.

Take a matrix element between states $\langle\chi|$ and
$U_2(\psi,\rho)|\overline\chi\rangle$ for
arbitrary $\psi\in\Hil(\Lambda)^0$ (note that $\overline\chi
\in\Hil_2^0$). Note that no information is lost by doing this
because of the irreducibility of the representations and the
fact that the intertwining and duality relations we already have
allow us to use the vertex operator corresponding to the state $\psi$
to raise the ground states to an arbitrary state
in the corresponding sectors. Moving the $U_2$ into the middle 
as a $V$ by
use of an intertwining relation and using the definition of $P$
gives us the relation
\begin{equation}
w^{-2\delta}\langle P(\chi,\chi;-z^\ast)|V(\psi,\rho)|P
\left(\overline\chi,
\overline\chi;-{1\over w}\right)\rangle=
z^{-2\delta}\langle P(\chi,\chi;-w^\ast)|V(\psi,\rho)|P
\left(\overline\chi,
\overline\chi;-{1\over z}\right)\rangle\,,
\end{equation}
or, more compactly,
\begin{equation}
\label{eqA}
G(\chi,\psi;\alpha,\beta)=\beta^{h_\psi-2\delta}G\left(
\chi,\psi;
\alpha\beta,{1\over\beta}\right)\,,
\end{equation}
for all $\psi\in\Hil(\Lambda)^0$,
where $h_\psi$ is the conformal weight of $\psi$ and
\begin{equation}
G(\chi,\psi;\alpha,\beta)\equiv\langle P(\chi,\chi;1)|V(\psi,\alpha)
|P(\overline\chi,\overline\chi;\beta)\rangle\,.
\end{equation}
[We use $x^{L_0}V(\psi,\rho)x^{-L_0}=x^{h_\psi}V(\psi,z\rho)$ and
$x^{-L_0}P(\chi,\chi;z)=P(\chi,\chi;xz)$.]
\item$\hat W(\chi,z)\chi=e^{zL_{-1}}\hat W(\chi,-z)\chi$.

This ``skew-symmetry'' relation (an analog of ($\ref{Wdef}$))
on the state $R(\chi,\chi;z)$ could be translated into a
corresponding requirement on the state $Q\in\Hil(\Lambda)^0$.
However, the philosophy which we are trying to pursue here
is to check everything in terms of matrix elements
of the form ($\ref{answer}$). So we choose to express
matrix elements of $R$ as the $d^{-\delta}$ term in
the expansion of ($\ref{answer}$) around $d=0$,
and hence find that our locality relation is equivalent
to the requirement that
\begin{equation}
\label{eqB}
O\left(F(\chi,\psi;\alpha,\beta),\beta,\delta\right)
=(-1)^{h_\psi+\delta}O\left(F(\chi,\psi;-\alpha-1,\beta),\beta,\delta\right)\,,
\end{equation}
where $O(H(\beta),\beta,\delta)$ denotes the $\beta^{-\delta}$ term in the
expansion of $H(\beta)$ around $\beta=0$ and
\begin{equation}
F(\chi,\psi;\alpha,\beta)\equiv\langle P(\chi,\chi;1|V(\psi,
\alpha)|P(\chi,\chi;\beta)\rangle\,.
\end{equation}
\item $\overline{W_2}(\chi,z)\hat W(\chi,w)=
\overline{W_2}(\chi,w)\hat W(\chi,z)$.

(Recall,
as remarked above,
in the notation of section 3, rather unfortunately,
$\overline{W_2}$ is $\overline{W_3}$!) Acting on the state
$\chi$ with this relation gives us a symmetry of the state
$Q(\chi,\chi;w,z)$, provided we have $\overline{\chi_3}=\chi$. There
are two steps to this. First, we must use the above result that $\chi_3$,
the
leading term in the expansion of $R(\chi,\chi;z)$ about $z=0$, lies
in the ground state of $\Hil_2^0$, {\em i.e.} check one of the
assumptions of the ansatz of \cite{PSMthird} by confirming
$\delta=d/9$. Secondly, we then simply choose a ``gauge'' in which
the spinor states are such that $\overline{\chi_3}=\chi$ by
suitable redefinition of $\chi$ by action of an appropriate unitary
matrix.

The locality requirement then becomes, from ($\ref{t22}$),
the requirement of symmetry of $Q\left(\chi,\chi;-{1\over z},
{zw\over z-w}\right)$ under $w\leftrightarrow z$.

As we have remarked above, however, we wish to translate everything
into requirements on the matrix elements given by the functions
$F$ and $G$ defined above.
We therefore use the identity ($\ref{21}$) and the relation between
($\ref{expm}$) and ($\ref{answer}$) to rewrite the locality relation as 
a requirement for symmetry under $w\leftrightarrow z$ of
the $d^{-\delta}$ term in the expansion about $d=0$ of
\begin{equation}
\langle P\left(\chi,\chi;\left({zw\over z-w}\right)^\ast\right)|
V\left(z^{2L_0}e^{-zL_1}\overline\psi,{z^2\over w-z}\right)
|P(\chi,\chi;d)\rangle\,,
\end{equation}
for all $\psi\in\Hil(\Lambda)^0$.
It clearly suffices to take $\psi$ quasi-primary.
Then, with some trivial manipulations, we see that we require,
for all quasi-primary $\psi\in\Hil(\Lambda)^0$ of conformal
weight $h_\psi$,
\begin{equation}
\label{eqC}
O\left(F\left(\chi,\psi;{1\over \alpha},\beta\right),
\beta,\delta\right)=(-1)^{h_\psi+\delta}\alpha^{2h_\psi}
O\left(F(\chi,\psi;\alpha,\beta),\beta,\delta\right)\,.
\end{equation}
\item $\overline{\hat W}(\overline\chi,z)\hat W(\chi,w)=
W_2(\chi,w)\overline{W_1}(\overline\chi,z)$.

Taking suitable matrix elements, as above, this is simply
(making a simple change of variables and writing the left hand side
in terms of $R$)
\begin{equation}
\langle R(\chi,\chi;d^\ast)|U_2(\psi,w)|R(\chi,\chi;z)\rangle=
\langle P(\chi,\chi;-z^\ast)|V(\psi,w)|P(\chi,\chi;-d)\rangle\,,
\end{equation}
or
\begin{equation}
\label{auto}
\langle R(\chi,\chi;d^\ast)|U_2(\psi,w)|R(\chi,\chi;z)\rangle=
(-z)^{-h_\psi}F\left(\chi,\psi;-{w\over z},dz\right)\,.
\end{equation}
\end{enumerate}

We have reduced verification of the consistency
of $\widetilde\Hil(\Lambda)$ to checking the four relations ($\ref{eqA}$),
($\ref{eqB}$),
($\ref{eqC}$) and ($\ref{auto}$) in terms of the functions $F$ and $G$. ($R$ is defined by $F$, as discussed previously, and in principle
($\ref{auto}$) can be checked -- we will comment further on this below
however.)

It should be stressed
that checking the consistency of the orbifold theory is in any
case a non-trivial requirement. For example,
in the case of a reflection-twisted
orbifold of the theory $\Hil(\Sigma)$ corresponding to an even lattice $\Sigma$
\cite{DGMtwisted}, the orbifold is only consistent for $\sqrt 2\Sigma^\ast$
an even lattice. A similar condition must hold in this case, though
presumably our restriction to consideration of an even self-dual lattice
renders the $\ze_3$-twisted orbifold theory consistent.

Some parts of the calculation of $F$ and $G$ were performed in
\cite{PSMthird}. We have, in the case $d=0$ mod 36 (so that $\chi$ is a pure
spinor ground state, {\em i.e.} $\chi\in{\cal G}$)
\begin{equation}
P(\chi,\chi;z)=\sum_{\lambda\in\Lambda}\langle\chi|\gamma_\lambda|\chi\rangle
3^{-{3\over 2}\lambda^2}z^{-\lambda^2}
e^{\sum_{{n,m\geq 0\atop n+m>0}}A_{mn}b_{-m}\cdot\overline b_{-n}z^{-m-n}}
|\lambda\rangle\,,
\end{equation}
(for some suitable ``cocycles'' $\gamma_\lambda$ acting on the
spinor ground state in which $\chi$ lies) where
\begin{equation}
A_{mn}={1\over m+n}
\left({-{1\over 3}\atop m}\right)
\left({-{2\over 3}\atop
n}\right)\,.
\end{equation}
$P(\overline\chi,\overline\chi;z)$ is given by the same expression
but with what turns out to be the
important difference that the matrix $A$ is transposed.
In evaluating the function $F$, we are faced with such expressions as
$\det (1-AA^T)$ (since ${b_n}^\dagger=\overline b_{-n}$),
whereas $G$ involves $\det (1-A^2)$. As remarked in \cite{PSMthird},
the former is problematic, and its solution could only be
conjectured, whereas the latter was calculated explicitly (in appendix C
of \cite{PSMthird}) by simply adapting the arguments of \cite{DGMtwisted},
which in turn follow from earlier calculations performed in the context
of dual models \cite{dualferm1,SchwarzWu:dualII}.

We see now from our observation ($\ref{Pred}$) (due to the
``reality'' of the representation) that $F$ can be evaluated
in terms of $G$ and much of the calculation proposed in
\cite{PSMthird} is unnecessary.

In fact, we have
\begin{equation}
\label{GFrel}
F(\chi,\psi;\alpha,\beta)=(1-\beta)^{-h_\psi}(-\beta)^{h_\psi}G\left(\chi,\psi;
{\alpha\beta+1\over\beta-1},1-\beta\right)\,,
\end{equation}
and we simply check the relations ($\ref{eqA}$),
($\ref{eqB}$), ($\ref{eqC}$) and
($\ref{auto}$) in terms of the single function $G$.
In fact, ($\ref{eqB}$) follows easily from ($\ref{eqA}$)
as a consequence of ($\ref{GFrel}$) (actually a  non-limiting
form of ($\ref{eqB}$), {\em i.e.} with the $O(\cdot)$ removed, holds).
This, together with the following important conjecture,
is a crucial simplification of the orbifold analysis.

We conjecture that ($\ref{auto}$) should hold as a consequence of our
definition of $R$.
First note that the identity between ($\ref{answer}$) and the
left hand side of ($\ref{auto}$) derived earlier is consistent with
($\ref{auto}$) if
\begin{equation}
(1-\beta)^{-2\delta} F\left(\chi,\psi;-1-\alpha,{\beta\over
\beta-1}\right)=(-1)^{h_\psi}F(\chi,\psi;\alpha,\beta)\,.
\end{equation}
Translating this into a requirement on $G$ via ($\ref{GFrel}$)
it is no more than ($\ref{eqA}$).
Now, the matrix elements of $R$ are defined by the leading term
in the expansion around $d=0$ of the right hand side of ($\ref{answer}$)
(or ($\ref{auto}$)), and hence so is the left hand side
of ($\ref{auto}$).
In other words, we should be able to deduce $F(\chi,\psi;\alpha,\beta)$
from the $\beta^{-\delta}$ term in the expansion about $\beta=0$.
This is certainly clear when $\delta=0$, since derivatives of $F$
with respect to $\beta$ are implemented by simple insertion of an appropriate
Virasoro mode, {\em i.e.} considering $\psi$ replaced by $V(\psi_L,z)\psi$.
The case when $\delta\neq 0$ is not so clear, and though intuitively we expect
this still to hold, many details remain to be checked.
Even then, the implication that ($\ref{auto}$) follows from this property
needs still to be established rigorously, and clearly ties in
with the still to be proven definition of a module in terms of the
state $P$ sketched in \cite{PSMreps}. This is the subject of further work.

[Some more weight may be lent to these conjectures by the following
observations. We note that the state $R(\chi,\chi;z)$ is itself
of the form of a ``P'' state, though of a more generalised form
since it lies in a non-adjoint module. It obeys similar equations,
(provided the orbifold is consistent) and by considering
the matrix element
\begin{equation}
\langle R(\chi,\chi;1)|e^{-L_1}U_2(\psi,w)|R(\chi,\chi;z)\rangle
\end{equation}
for $\psi$ quasi-primary
and acting with the exponential to the right or to the left \cite{thesis}
we can simply deduce the identity
\begin{equation}
\langle R(\chi,\chi;-1)|U_2(\psi,w)|R(\chi,\chi;z)\rangle=
(1+w)^{-2h_\psi}(1+z)^{-2\delta}
\langle R(\chi,\chi;1)|
U_2\left(\psi,{w\over w+1}\right)|R\left(\chi,\chi;{z\over z+1}\right)
\rangle\,.
\end{equation}
This is a potential ``hidden'' symmetry, which must lead to a symmetry
of $F$ and hence $G$ via ($\ref{auto}$). If such an identity were non-trivial
it would be very mysterious and cast serious doubt on any
claims that ($\ref{auto}$) is a simple consequence of the definition
of $R$ -- since it would then follow from properties of $R$
in the twisted module that were not apparent at the level of
the untwisted sector. Fortunately, however, the symmetry
one derives is again the identity ($\ref{eqA}$).]

Modulo these assumptions then, we see that verification of
locality is reduced to simply checking the two
relations ($\ref{eqA}$)
and ($\ref{eqC}$) inside $\Hil(\Lambda)^0$.
We stress once more that, in keeping with our philosophy
throughout this paper, no calculations in the twisted sectors
are necessary (and even if we were to consider
($\ref{auto}$) explicitly, the structure of the module in
which $R$ lies is induced by the matrix elements given
by $F$, as discussed in the last section).

Note also that no assumptions on the form of the intertwiner
between the twisted sectors need be made. Indeed, we may check the
ansatz given in \cite{PSMthird} (for $d$ a multiple of 36)
\begin{equation}
R(\chi,\chi;z)=z^{-d/9}e^{\sum_{r\in\ze+{1\over 3},
s\in\ze+{2\over 3}}D_{rs}\overline c_{-r}\cdot\overline c_{-s}(-z)^{
r+s}}\chi_3\,,
\end{equation}
where
\begin{equation}
D_{rs}={1\over r+s}\left({1\over 3s}-1\right)
\left({-{2\over 3}\atop r-{1\over 3}}\right)
\left({-{1\over 3}\atop s-{2\over 3}}\right)\,,
\end{equation}
(for some ground state $\chi_3\in\Hil_2^0$),
by use of our relation
\begin{eqnarray}
\langle\chi_3|U_2(\psi,w)|R(\chi,\chi;z)\rangle&=&O\left(z^{-h_\psi}
z^{-\delta}F(\chi,\psi;{w\over z}-1,d);d,\delta\right)\nonumber\\
&=&O\left(z^{-h_\psi}
z^{-\delta}(-d)^{h_\psi}G\left(\chi,\psi;{wd\over z(1-d)}+1,1-d\right);
d,\delta\right)\,,
\end{eqnarray}
(where $\delta$ is given by the order of the leading pole in $F$).

For $d\equiv 12$, 24 mod 36, the correct from of $R(\chi,\chi;z)$
was unclear in \cite{PSMthird} and an appropriate ansatz could not be made.
Our present reformulation sidesteps this problem by allowing one
to calculate $R$ if desired, but moreover arguing
that
such expressions are unnecessary in any case and the appropriate $P$
(different for $d\equiv 12$, 24 mod 36 from that given above,
though still easily calculated) is all that is required.

The calculations remaining to check consistency of $\widetilde\Hil(\Lambda)$,
{\em i.e.} evaluating $G(\chi,\psi;\alpha,\beta)$ for all $\psi\in\Hil_2^0$,
are just a straightforward
generalisation of those performed in \cite{DGMtwisted}. However,
they are clearly outside the scope of this short paper and will
be presented in a future publication.

Clearly, the ideas here extend beyond the simple $\ze_3$ case which
we have used as an illustrative example. We may even analyse
orbifolds of arbitrary theories where no explicit realisation of the twisted
sectors is apparent,
perhaps because, for example, the orbifold does not have
the same geometric interpretation as is available in the $\ze_3$ case.
\section{Conclusions}
Though the main idea of this paper is, in some sense, a rather trivial rewriting of
matrix elements involving intertwiners, the same is true of the
earlier work in \cite{PSMreps,PSMtwisunique} and yet such work led to
highly non-trivial results on the uniqueness of certain twisted
representations and very significant conjectures regarding ``induced''
representations of a conformal field theory from a sub-conformal field
theory contained within. Similarly in this paper, we have shown that,
in the context of orbifold theories endowed with a hermitian
structure, the intertwiners between twisted sectors are unique, as one
would clearly expect but a conjecture which had
up to now remained unproven in general, and moreover we have given a means of
actually
calculating matrix elements involving these intertwiners in terms of
the starting representations. As before, all calculations are carried
out entirely within the underlying (twist-invariant, in the case of
orbifolds) conformal field theory, and no explicit knowledge of the
twisted sectors need be used, or indeed known.
A proposed technique for verifying consistency of arbitrary (non-geometric)
orbifolds has been laid out.

Also, one can view this technique as a means of generating new
representations (in the fusion product) from given ones.
Our
calculations produce the state $P(\chi_3,\chi_3;z)$ from which we may
construct the representation $\kay_3$ as in \cite{PSMreps}.  This
obviously leads us to consider the case in which fusion of the two
representations $\kay_1$ and $\kay_2$ gives rise to more than one
irreducible representation (note that we have implicitly appealed to
Dong {\em et al}'s proof of the uniqueness of the twisted
representation \cite{DLMZhu} in the discussion of this paper, {\em
i.e.} we assume that in fusing a $g$-twisted representation with an
$h$-twisted representation the resulting $gh$-twisted representation
is unique). Note that ($\ref{answer}$) is, by definition,
simply $\langle P(\hat W(\chi_1,d^\ast)\chi_2,
\hat W(\chi_1,w,\chi_2);w)|\psi\rangle$ (we may
easily recast this into an explicit form for $P$,
{\em i.e.} not involving inner products, by
use of a relation for the vertex operators
$V$ analogous to ($\ref{Wdef}$)), constructed solely from
the given representations
$\kay_1$ and $\kay_2$. If their fusion
involves more than one irreducible representation, this state $P$ will
split into a sum of $P$'s corresponding to each of these. It would be
expected then that a projection onto states orthogonal
to $O(\Hil)$ \cite{Zhu} (or equivalently taking inner products with respect to
the $P$'s corresponding to the irreducible representations, if known --
{\em i.e.} considering matrix elements
$\langle P_i(y)|V(P_j,w)|P_k(z)\rangle$)
would enable one to deduce the fusion rules. However, this procedure
is complicated by the fact that the states appearing
in $\hat W(\chi_1,z)\chi_2$ do not necessarily
include ground states in all of the representations
which occur, but rather some representations may
appear starting at a higher level.
The relation to Frenkel and Zhu's calculations of \cite{FrenkelZhu} remains to be
made clear.
This and related questions form the basis of current research.
\appendix
\section{Failure of the naive ansatz for the intertwiner}
We demonstrate in this appendix that an obvious but naive ansatz
for the form of the intertwiner representing the action of $\kay_2$
on $\kay_3$ fails in general. In the process, we derive some very
interesting results related to Zhu's algebra. As yet, we do not yet
fully understand the significance of these.

Throughout this appendix, $\psi$ will denote a state in $\Hil$
of conformal weight $h$.
Choose $\chi_1$ and $\chi_2$ to be ground states in $\kay_1$ and
$\kay_2$ respectively. Then, using the
result
\begin{equation}
\label{identity}
\sum_{n=0}^h\left({h\atop n}\right)(-1)^n\left({h+n\atop n+s}\right)=0\,,
\end{equation}
for $s=1,\ldots,h$
(which is easily checked by induction), we can easily show from
($\ref{duality}$) that
\begin{equation}
\left[\sum_{n=0}^h\left({h\atop n}\right)(-1)^nV(\psi)_{n+1},V(\phi,z)\right]
=0\,,
\end{equation}
for $\phi$ a highest weight state under the action of $\psi$, {\em i.e.}
$V(\psi)_n\phi=0$ for $n>0$.

In particular, applying this in the orbifold conformal field theory, we
have
\begin{equation}
\sum_{n=0}^h\left({h\atop n}\right)(-1)^nV(\psi)_{n+1}\hat W(\chi_1,z)
\chi_2=0\,.
\label{sum}
\end{equation}
Now we make the ``natural'' ansatz
\begin{equation}
\hat W(\chi_1,z)\chi_2=U_3\left(H(z),z)\right)\chi_3\,,
\end{equation}
with $\chi_3$ as in section \ref{Q} (except we now assume it lies
in the ground state of $\kay_3$) and $H(z)$ a state in $\Hil$ to be
determined.
Duality calculations together with ($\ref{sum}$) then imply
\begin{equation}
\sum_{n=0}^h\left({h\atop n}\right)(-1)^n\sum_{s=-n}^h\left({h+n\atop s+n
}\right)V(\psi)_{-s+1}H(1)=0\,.
\end{equation}
An extension of ($\ref{identity}$), namely
\begin{equation}
\sum_{n=0}^h\left({h\atop n}\right)(-1)^n\left({h+n\atop n-r}\right)=-
\left({h\atop r}\right)\,,
\end{equation}
gives
\begin{equation}
\sum_{r=0}^h\left({h\atop r}\right)V(\psi)_{r+1}H(1)=0\,.
\end{equation}

Let us briefly recall Zhu's construction of $O(\Hil)$ \cite{Zhu},
a subspace of $\Hil$ whose zero modes annihilate the ground state
of any representation of $\Hil$. $O(\Hil)$ is constructed as the span of
states $O(\psi,\phi)$ for all pairs $\psi$, $\phi\in\Hil$, where
\begin{equation}
O(\psi,\phi)=\sum_{r=0}^h\left({h\atop r}\right)V(\psi)_{-r-1}\phi\,.
\end{equation}
Now, ($\ref{conj}$) implies that ${V(\psi)_r}^\dagger=V
\left(e^{L_1}\overline\psi\right)_{-r}$, and so we see that $H(1)$
is orthogonal to the subspace of $O(\Hil)$ spanned by $O(\psi,\phi)$
for all pairs $\psi$, $\phi\in\Hil$ with $\psi$ quasi-primary.

We quote a very useful lemma (we omit the trivial proof):

\leftline{\bf Lemma} $O(L_{-1}\psi,\phi)=-hO(\psi,\phi)+(L_0+L_{-1})
O(\psi,\phi)-O(\psi,(L_0+L_{-1})\phi)$.

In addition, we can show, using the commutation relations of $su(1,1)$
with vertex operators detailed in {\em e.g.} \cite{PGmer},
\begin{equation}
\label{relnL}
(L_1+L_0)H(1)=(h_1+h_3-h_2)H(1)\,.
\end{equation}
But $(L_1+L_0)H(1)=\kappa H(1)$ for some scalar $\kappa$,
together with the lemma and the fact that $H(1)$ is
orthogonal to $O(\psi,\phi)$ for all $\psi$ quasi-primary
implies that $H(1)$ is orthogonal to $O(\Hil)$
by induction.

Now, $(L_{-1}+L_0)\phi\in O(\Hil)$ for all $\phi\in\Hil$. Hence, we must
have
\begin{equation}
(L_1+L_0)H(1)=0\,,
\end{equation}
contradicting ($\ref{relnL}$) unless $h_1+h_3=h_2$. Typically this is
not the case, and so we deduce that the ansatz for the form of $\hat
W(\chi_1,z)\chi_2$ is inconsistent in general.  For example, in a
$d$-dimensional lattice theory twisted by a third order automorphism
$g$ induced by a no-fixed-point third order automorphism of the
lattice the two twisted sectors both have ground states of conformal
weight ${d\over 18}$
(take $d$ to be a multiple of 72) \cite{PSMthird}.
It happens that when $\chi_1$ and $\chi_2$ are
taken to be identical states in the ground state of the $g$-twisted
sector, then the corresponding state $\chi_3$ lies in the ground state
of the $g^2$-twisted sector \cite{PSMthird}. In this case, our ansatz
for the form of $\hat W(\chi_1,z)\chi_2$ must fail.

However, there are examples in which the ansatz can, and does, hold.
For example, consider representations of the $d$-dimensional
Heisenberg algebra.
These are described by a $d$-dimensional
``momentum'' vector $\lambda$
corresponding to the
ground state of the representation. The corresponding conformal weight
is given by $h=\lambda^2/2$. In fusing the representations, the momenta
simply add. Hence our ansatz can only make sense for the action
of a representation labelled by $\lambda$ on one labelled by $\mu$
if $\lambda^2+\lambda\cdot\mu=0$. In one dimension, this is just
the statement that $\lambda=-\mu$ and the target representation is simply
the adjoint. In fact in this case $\hat W(\chi_\lambda,z)\chi_\mu=
z^{-\lambda^2}P(\chi_\lambda,\chi_\mu;z)$, trivially from the
definition ($\ref{Pdef}$).
$\chi_3$ is then simply the vacuum $|0\rangle$, and ($\ref{create}$)
tells us that
\begin{equation}
H(z)
=e^{-zL_{-1}}z^{-\lambda^2}P(\chi_\lambda,\chi_\mu;z)\,.
\end{equation}

\end{document}